# A Study of the Stability Properties of Smooth Particle Hydrodynamics

**Joseph Peter Morris,**
Mathematics Department, Monash University, Clayton VIC 3168
email: jpm@fermat.maths.monash.edu.au
(To appear in the Proceedings of the Astronomical Society of Australia)

**Abstract:** When using a formulation of Smooth Particle Hydrodynamics (SPH) which conserves momentum exactly the motion of the particles is observed to be unstable to negative stress. It is also found that under normal circumstances a lattice of SPH particles is potentially unstable to transverse waves. This document is a summary of a detailed report (Morris 1994) investigating the nature of these and other instabilities in depth. Approaches which may be used to eliminate these instabilities are suggested. It is found that the stability properties of SPH in general improve as higher order spline interpolants, approximating a Gaussian, are used as kernels.

## 1. Introduction

The early applications of SPH, Lucy 1977 and Gingold and Monaghan 1977, were to problems involving a compressible gas which always had a positive gas pressure. The equations governing the motion of the particles, when written in a form which conserves momentum exactly, result in particles repelling each other with equal and opposite forces. As particles approach, the density increases, the pressure increases, and the particles tend to repel each other. When applied to different problems where the stress can become negative the momentum conserving form of SPH is observed to become unstable to short wavelength perturbations. For negative stress, the particles no longer repel, but attract. Each particle is, in effect, at the bottom of a potential well, and it becomes possible for the particles to pair up and "slide" into each others wells, causing "clumping" initially and subsequently disrupting the solution. The nature of this instability is quite different from instabilities observed in explicit finite difference techniques when too large a time step is used. This SPH instability is a consequence of the particles being free to move under negative stress and is present even if the time integration is exact. Additional stability conditions, such as the Courant Friedrichs Levy condition must also be respected. Recently people have commented upon this instability in connection with many problems (e.g. elastic flow in Swegle 1992), but much earlier Phillips and Monaghan 1985 had detected and studied the problem in connection with magnetohydrodynamics (MHD). Very little analytical work has been done in the study of the stability properties of SPH and previous methods (see Monaghan 1989) actually break down in the region where the stability we seek to understand first appears. The report (Morris 1994) extends the stability analysis of SPH in order to not only understand the nature of potential instabilities but to give insight into how methods giving most accurate results may be formulated.

## 2. Seeking the Source of the Instability

In order to understand the problem better we must seek the fundamental source of instability. We have seen that the reported instability occurs when a stress tensor is implemented in a form which conserves momentum exactly. We will be considering one dimensional flows initially, with symmetric kernels. Typically, we might use a Gaussian kernel,

$$W(\mathbf{x}, h) = \frac{1}{h\pi^{\frac{1}{2}}} \exp\left(-\left\{\frac{|\mathbf{x}|}{h}\right\}^2\right), \quad (1)$$

or the cubic spline interpolant,

$$W(\mathbf{x}, h) = \frac{1}{h} \begin{cases} \frac{2}{3} - \left(\frac{|\mathbf{x}|}{h}\right)^2 + \frac{1}{2}\left(\frac{|\mathbf{x}|}{h}\right)^3, & \text{if } 0 \le |\mathbf{x}| \le h; \\ \frac{1}{6}\left(2 - \left(\frac{|\mathbf{x}|}{h}\right)\right)^3, & \text{if } h \le |\mathbf{x}| \le 2h; \\ 0, & \text{if } |\mathbf{x}| \ge 2h. \end{cases} \quad (2)$$

Choosing a symmetric kernel guarantees exact conservation of momentum, since inter-particle forces then form action-reaction pairs.

Let us consider the specific case of one-dimensional MHD flow, with a constant magnetic induction $B_0$ parallel to the direction of the flow, say along the $x$-axis. Since the magnetic induction is constant, the equations of MHD reduce to those for hydrodynamics. However, the corresponding SPH equations may be written,

$$\begin{aligned}
\frac{dv_a}{dt} &= -\sum_b m_b \left(\frac{p_a - \frac{1}{2\mu_0}B_0^2}{\rho_a^2} + \frac{p_b - \frac{1}{2\mu_0}B_0^2}{\rho_b^2}\right) \frac{\partial W_{ab}}{\partial x_a}, \\
\rho_a &= \sum_b m_b W_{ab}, \quad p_a = c^2 \rho_a, \\
W_{ab} &= W(x_a - x_b, h).
\end{aligned} \quad (3)$$

So, in this case, the SPH equations for magnetohydrodynamical equations are equivalent to those for pure hydrodynamical flow with the pressure adjusted by a constant. Similar analysis may be done for any problem involving a stress tensor of similar form. So, the stability properties of one-dimensional Smooth Particle Magnetohydrodynamics (SPMHD) will be exhibited by the equations,

$$\begin{aligned}
\frac{dv_a}{dt} &= -\sum_b m_b \left(\frac{p_a}{\rho_a^2} + \frac{p_b}{\rho_b^2}\right) \frac{\partial W_{ab}}{\partial x_a}, \\
\rho_a &= \sum_b m_b W_{ab}, \quad p_a = c^2 \rho_a + P,
\end{aligned} \quad (4)$$

where $P = -B_0^2/2\mu_0$. Similar equations, with a different expression for $P$ will be obtained by considering the one-dimensional case of other applications of SPH.

## 3. Stability Analysis and SPH

A simple way to analyse the stability of an implementation of SPH is to consider the dispersion relation for linear, sound waves propagating in a one-dimensional flow. We can imagine an infinite line of identical particles oscillating about a constant mean separation. Such a wave may be described by,

$$x_a = a\Delta x + X \exp(ika\Delta x - i\omega t), \quad (5)$$

We now linearise Eqs. (4) by substituting Eq. (5) and neglecting all but first order terms to obtain,

$$\begin{aligned}
\omega^2 &= \frac{2mc^2\Re}{\rho_0} \sum_j (1 - \cos k\Delta xj) \frac{\partial^2 W}{\partial x^2}(\Delta xj, h) \\
&\quad + \left(\frac{mc}{\rho_0}\right)^2 (1 - 2\Re) \left\{\sum_j \sin k\Delta xj \frac{\partial W}{\partial x}(\Delta xj, h)\right\}^2,
\end{aligned} \quad (6)$$

where we have chosen, $P = c^2 \rho_0 (\Re - 1)$. I have assumed $W$ to be even (a more general expression is given in Morris 1994). This choice of $P$ gives,

$$p_a = c^2 \rho_0 \left(\frac{\rho_a}{\rho_0} - 1 + \Re\right). \quad (7)$$

We can think of the quantity $c^2 \rho_0 \Re$ as being the "background" pressure upon which the linear wave perturbations occur. Thus, if $\Re > 0$, then $p_a$ should be positive for small perturbations and if $\Re < 0$, then $p_a$ may become negative. $\Re$ will take different values depending on the equation of state being considered. For standard SPH, $\Re = 1$. For SPMHD, $\Re = 1 - B_0^2/2\mu_0 c^2 \rho_0$. For incompressible SPH (see Monaghan 1994), $\Re = 0$. Using Poisson's summation formula, Eq. (6) may be written,

$$\omega^2 = \frac{2mc^2\Re}{\rho_0 \Delta x} \sum_{l=-\infty}^{+\infty} \left\{ \left(k + \frac{2\pi l}{\Delta x}\right)^2 U\left(k + \frac{2\pi l}{\Delta x}, h\right) \right.$$
$$\left. - \left(\frac{2\pi l}{\Delta x}\right)^2 U\left(\frac{2\pi l}{\Delta x}, h\right) \right\}$$
$$+ \left(\frac{mc}{\rho_0 \Delta x}\right)^2 (1 - 2\Re) \left\{ \sum_{l=-\infty}^{+\infty} \left(k + \frac{2\pi l}{\Delta x}\right) U\left(k + \frac{2\pi l}{\Delta x}, h\right) \right\}^2 \quad (8)$$

### 4. Stability of Another Formulation

It is possible to develop an alternative formulation to Eq. (4) which has stability properties independent of the background pressure,

$$\frac{dv_a}{dt} = -\sum_b m_b \frac{p_b - p_a}{\rho_a \rho_b} \frac{\partial W_{ab}}{\partial x_a}. \quad (9)$$

Since the differences of the pressures are taken in the momentum equation, Eq. (9), the value of $\Re$ is irrelevant. However, the forces exerted upon the particles according to Eq. (9) no longer form action-reaction pairs but are estimates of the local pressure gradient. Thus, this formulation does not conserve momentum exactly, except in the continuum limit (ie: $h \to 0$ and infinitely many particles). This pressure difference formulation has a very simple equation governing its stability,

$$\omega^2 = \left(\frac{mc}{\rho_0}\right)^2 \left[ \sum_{j=-\infty}^{+\infty} \sin k\Delta x j \frac{\partial W}{\partial x}(\Delta x j, h) \right]^2. \quad (10)$$

### 5. The Numerical Sound Speed

Ideally, we have $\omega^2 = c^2 k^2$, thus it is useful to define,

$$C_{num}^2 = \frac{c_{num}^2}{c^2} = \frac{\omega^2}{k^2 c^2},$$

where $c$ is the analytic sound speed. Ideally, of course, $C_{num}$ should be as close to one as possible, so that sound waves are propagated realistically. If $C_{num}^2$ becomes negative, the perturbations in the numerical solution are no longer travelling waves, but are exponentially growing and decaying disturbances.

### 6. Results

It is clear that the dispersion relation is periodic in $k$ with period $2\pi/\Delta x$. This is the result of aliasing, whereby waves of wavenumber $k$ and $k + 2\pi/\Delta x$ look identical at a series of points separated by $\Delta x$. It has been established experimentally that the "problem" wavenumber is $\pi/\Delta x$ (a wavelength of two particle spacings). For $k = \pi/\Delta x$, Eq. (8), becomes,

$$\omega^2 = \frac{2mc^2\Re}{\rho_0 \Delta x} \sum_{l=-\infty}^{+\infty} \left\{ \left(\frac{\pi(2l+1)}{\Delta x}\right)^2 U\left(\frac{\pi(2l+1)}{\Delta x}, h\right) \right.$$
$$\left. - \left(\frac{2\pi l}{\Delta x}\right)^2 U\left(\frac{2\pi l}{\Delta x}, h\right) \right\},$$
$$= \frac{4mc^2\Re}{\rho_0} \sum_{j=-\infty}^{\infty} \frac{\partial^2 W}{\partial x^2}(\Delta x[2j+1], h). \quad (11)$$

Eq. (11) shows that, for wavenumber $\pi/\Delta x$, the sign of $\omega^2$, and hence $C_{num}^2$, changes sign with $\Re$. Thus, for any symmetric kernel, as the background pressure $c^2 \rho_0 \Re$ changes sign, it will change stability. There can be no kernel which is stable for both signs of $\Re$, using this implementation of SPH. Thus, to achieve general stability for different background pressures, we must modify the kernel as $\Re$ changes sign. Let us consider Eq. (11) for the simple case where $W(\Delta xj, h) = 0$ for $|j| > 2$. Then,

$$\omega^2 = \frac{8mc^2\Re}{\rho_0} \frac{\partial^2 W}{\partial x^2}(\Delta x, h). \quad (12)$$

So, for this case, the sign of $\partial^2 W/\partial x^2(\Delta x, h)$ must change when the background pressure changes sign. This is particularly important to remember if we are seeking a kernel which changes continuously with $\Re$ to provide stability. The appropriate kernels for $\Re$ being slightly positive and slightly negative have different signs for $\partial^2 W/\partial x^2$, so $\partial^2 W/\partial x^2$ must be zero at the nearest neighbour when $\Re = 0$ for continuity. In the case of the Gaussian, we find that $\omega^2 = 0$ when $h = 2^{\frac{1}{2}}\Delta x$. Thus, varying the smoothing length may be sufficient to ensure stability for this wavenumber. A useful solution, however, must be stable for all wavenumbers.

We can see from the form of Eq. (6) that for any given $h$ and $k$, $C_{num}^2$ varies linearly with $\Re$. In general we wish $C_{num}^2$ to vary only slowly with $\Re$, since, for the exact system, it is independent of $\Re$. It is informative to consider the quantity $\partial C_{num}^2/\partial \Re$, since this reflects the degree of dependence of the stability properties on $\Re$. For sufficiently large $\Re$ we have that $C_{num}^2 \propto \Re$ where the constant of proportionality (for a given $k$ and $h$) is $\partial C_{num}^2/\partial \Re$.

The limit of Eq. (8) as $k$ approaches zero is of particular significance. It is in this limit that we expect our numerical method will resolve the wave best, since it corresponds to an infinite number of particles fitting into each wavelength. It can be shown that, in this limit, $C_{num}^2$ is perturbed from 1 by a term which is linear in $\Re$ and proportional to the Fourier transform of the kernel evaluated at $2\pi/\Delta x$. The Fourier transform of the Gaussian kernel is, of course a Gaussian in $kh$, and, accordingly, falls off quite rapidly with $kh$. The Fourier transforms of kernels with compact support, however, do not fall off so rapidly, and we expect the sound speed to depend more strongly upon $\Re$ for such kernels. Numerical results obtained using direct summation in Morris 1994 of Eq. (6) confirm that $C_{num}$ depends more strongly on $\Re$ for spline interpolated kernels. In fact, there are choices of $k$, $\Delta x$ and $h$ for which $\partial C_{num}^2/\partial \Re$ is negative. Thus, a sufficiently large choice of $\Re$ for such parameters results in $C_{num}^2$ becoming negative, producing an instability. This result was not anticipated since it had been thought that SPH would be stable to any positive stresses. This instability resulting from over-pressure is of long wavelength and, correspondingly, can have a serious affect upon the accuracy of large scale phenomena. The use of smoother spline approximations to a Gaussian kernel increases the over-pressure required to induce the instability. More details on these results may be found in Morris 1994.

### 7. Mutating the Kernel

We have seen that no single kernel can provide us with the stability we seek. We need some way of modifying the present Gaussian shaped kernel. Any kernel, $W(x, h)$ we choose must have the following properties (in one-dimension):

$$\int_{-\infty}^{+\infty} W(x,h) dx = 1, \quad \lim_{h \to 0} W(x,h) = \delta(x), \quad (13)$$

and, if we wish momentum to be conserved exactly,

$$W(x, h) = W(-x, h). \quad (14)$$

Suppose that our "favourite" kernel is,

$$W_0(x, h) = \frac{1}{h} f\left(\frac{x}{h}\right), \quad (15)$$

We can now define an infinite series of symmetric kernels satisfying Eqs. (13) of the following form,

$$W_n(x, h) = \frac{A_n}{h} \left(\frac{x}{h}\right)^{2n} f\left(\frac{x}{h}\right), \quad (16)$$

where $A_n$ is chosen to satisfy Eq. (13). Now we have a set of kernels which we can combine to form a kernel with the necessary properties for stability,

$$W(x,h) = \sum_{j=0}^{N} B_n W_n(x,h), \qquad (17)$$

where,

$$\sum_{j=0}^{N} B_n = 1. \qquad (18)$$

We need to develop some method for choosing the co-efficients $B_n$ such that the resulting kernel has the desired stability properties. One possible approach is to choose several points that we want $\omega(k)$ to pass through. We then solve for the same number of free co-efficients as we have conditions by linearising about an initial guess and solving the resultant linear system of equations. We continue this iterative process until sufficient convergence is achieved. There are some conditions which must be respected in order that the system converges. These, and other details, are explored in Morris 1994.

### 8. The Pressure Differencing Formulation of SPH

The form of Eq. (10) ensures that the formulation Eq. (9) is always stable. It can also be shown that the sound speed for this implementation depends less strongly upon the particle spacing and smoothing length than the exactly momentum conserving formulation. This implementation, of course, has stability independent of $\Re$ since all pressures appear as differences in Eq. (9). However, it does not conserve momentum exactly and, thus, is not as good as the momentum conserving formulation (see Steinmetz and Mueller 1993) for modelling strong shocks. The pressure differencing formulation exhibits much larger oscillations when modelling strong shocks, though these might be tamed by using a different form of viscosity. It is possible to devise a shock switch to detect and track shocks such that this viscosity is only present at the shock front. The exactly momentum conserving method calculates the net particle forces as the sum of individual repelling interactions. This is less subject to oscillations than a method where the forces exerted on the particles are calculated from an estimate of the pressure gradient. When modelling stronger shocks, the pressure differencing formulation becomes less accurate as exact conservation of momentum becomes more necessary.

### 9. Stability Analysis with Variable Particle Spacing

In practice, a simulation will not involve particles oscillating about a constant separation. In order to deal with a number of different particle spacings we could solve for a kernel appropriate to each particle spacing and then use symmetric combinations of these when particles interact. If the expression Eq. (6) is used to find appropriate kernels, it is found that the numerical sound speed is significantly reduced. Effectively, the kernel is being changed continuously in such a fashion as to slightly oppose the propagation of the wave. If, however, we reformulate the stability analysis to include the assumption that the kernel is a function of the particle spacing, the sound speed will be more accurate. We need an appropriate variable to reflect the local spacing of particles. The distance to a nearest neighbour and other similar quantities are quite clumsy to use for such analysis. The specific volume,

$$V_a = \frac{m}{\rho_a}, \qquad (19)$$

is much more suitable. We now consider Eqs. (4) with the consideration that,

$$W_{ab} = W(x_{ab}, h, V_{ab}), V_{ab} = \frac{1}{2}(V_a + V_b). \qquad (20)$$

Applying stability analysis to the resulting SPH equations, allows us to obtain a somewhat complicated dispersion relation. An appropriate form for the kernel can be derived in a similar fashion to that employed in the case of constant particle separation.

### 10. Stability Analysis with Viscosity

One might think that introducing viscosity to the equations of motion might eliminate the instability resulting from negative stress. In particular, since the first instability is of short wavelength ($k \approx \pi/\Delta x$) we would expect viscosity to have a strong damping effect upon it. Let us consider the previous equations with the addition of artificial viscosity,

$$\frac{dv_a}{dt} = -\sum_b m_b \left( \frac{p_a}{\rho_a^2} + \frac{p_b}{\rho_b^2} + \Pi_{ab} \right) \frac{\partial W_{ab}}{\partial x_a},$$
$$\Pi_{ab} = -\frac{\alpha c \mu_{ab}}{\frac{1}{2}(\rho_a + \rho_b)}, \mu_{ab} = \frac{h v_{ab} x_{ab}}{x_{ab}^2 + \eta^2}. \qquad (21)$$

Once again, we consider a linear wave perturbing an infinite line of particles parallel to the $x$-axis and obtain the following dispersion relation,

$$\omega^2 + ia\omega - b = 0, \qquad (22)$$

where,

$$a = -\frac{m\alpha ch}{\rho_0 \Delta x} \sum_{j\backslash\{0\}} \frac{1 - \cos k \Delta x j}{j} \frac{\partial W}{\partial x}, \qquad (23)$$

and

$$b = \frac{2mc^2\Re}{\rho_0} \sum_j (1 - \cos k\Delta x j) \frac{\partial^2 W}{\partial x^2}$$
$$+ \left( \frac{mc}{\rho_0} \right)^2 (1 - 2\Re) \left[ \sum_j \sin k\Delta x j \frac{\partial W}{\partial x} \right]^2. \qquad (24)$$

The definitions of $a$ and $b$ are chosen such that they are real and, for the standard kernel and $\Re = 1$, are positive. Thus, we have,

$$\omega = \pm \frac{(4b - a^2)^{\frac{1}{2}}}{2} - i\frac{a}{2}. \qquad (25)$$

From Eq. (5) we find,

$$v = V \exp(ikx - i\omega t) = V \exp \left( ikx \mp i\frac{(4b-a^2)^{\frac{1}{2}}}{2} t - \frac{a}{2} t \right). \qquad (26)$$

We see that $a$ is responsible for dampening the motion and reducing the speed of propagation. Let us now consider the crucial wavenumber, $k = \pi/\Delta x$. Substituting into Eqs. (23) and (24) we obtain,

$$a = -\frac{m\alpha ch}{\rho_0 \Delta x} \sum_{j\backslash\{0\}} \frac{1 - (-1)^j}{j} \frac{\partial W}{\partial x},$$
$$b = \frac{2mc^2\Re}{\rho_0} \sum_j (1 - (-1)^j) \frac{\partial^2 W}{\partial x^2}. \qquad (27)$$

Typically if $\Re < 0$ (ie: negative stress) then $b < 0$. Thus, we obtain,

$$\omega = \pm i \frac{(4\beta + a^2)^{\frac{1}{2}}}{2} - i\frac{a}{2}, \qquad (28)$$

where $\beta = -b$. The component of this which leads to the instability is,

$$\omega = \frac{i}{2} \left( (4\beta + a^2)^{\frac{1}{2}} - a \right), \qquad (29)$$

So we see that the instability is not removed by the introduction of viscosity. If $a^2/4\beta \ll 1$ then

$$\omega \approx i \left( \beta^{\frac{1}{2}} - \left( \frac{a}{2} - \frac{a^2}{8\beta} \right) \right). \qquad (30)$$

If $a/2 > a^2/8\beta$, then we see that the effect of introducing a small amount of viscosity is to reduce the growth rate of the instability.

## 11. Two Dimensional Stability Analysis

Let's consider the stability properties of standard two dimensional isothermal SPH,

$$\frac{d\mathbf{v}_{ab}}{dt} = -\sum_c \sum_d m_{cd}\left(\frac{p_{ab}}{\rho_{ab}^2} + \frac{p_{cd}}{\rho_{cd}^2}\right)\nabla_{ab}W_{abcd}$$
$$p_{ab} = c^2\rho_{ab}, \quad \rho_{ab} = \sum_c \sum_d m_{cd}W_{abcd} \quad (31)$$
$$W_{abcd} = W(x_{ab} - x_{cd}, y_{ab} - y_{cd}, h)$$

We can analyse the stability of a rectangular lattice of particles by considering the propagation of plane waves on such a grid,

$$x_{ab} = a\Delta x + X\exp(ik_x a\Delta x + ik_y b\Delta y - i\omega t),$$
$$y_{ab} = b\Delta y + Y\exp(ik_x a\Delta x + ik_y b\Delta y - i\omega t). \quad (32)$$

Proceeding in a similar fashion to the one dimensional analysis we obtain two solutions for each pair of $k_x$ and $k_y$. The full details are in Morris 1994, but let us consider the dispersion relation obtained for plane waves propagating along the $x$-axis. We find that there is a longitudinal wave solution,

$$\omega^2 = \frac{mc^2}{\rho_0}2\sum_i\sum_j(1 - \cos k\Delta xi)\frac{\partial^2 W}{\partial x^2}$$
$$- \left(\frac{mc}{\rho_0}\right)^2\left\{\sum_i\sum_j \sin k\Delta xi \frac{\partial W}{\partial x}\right\}^2, \quad (33)$$

which corresponds to the one dimensional result Eq. (6) with $\Re = 1$ and the kernel replaced by a sum over $j$ of the kernel. It turns out that this finite approximation to the $y$ integration of the kernel introduces some small instabilities for large $h/\Delta x$. The system also supports a transverse wave with the dispersion relation,

$$\omega^2 = \frac{mc^2}{\rho_0}2\sum_i\sum_j(1 - \cos k\Delta xi)\frac{\partial^2 W}{\partial y^2}. \quad (34)$$

These waves, of course, are not present in the exact solution to the wave equation for an isothermal gas, but they are present in our lattice approximating an isothermal gas. Their presence is the result of variations in the value of the gradient of the kernel as particles "jostle" around. It turns out (see Morris 1994) that for spline interpolated kernels $\omega^2$ is negative for about half of the choices of $h$ for a given $\Delta x$. The growth rate of these instabilities is radically reduced as higher order spline interpolants are used. If the Gaussian is employed the growth rate of these unstable transverse modes is negligible for all practical choices of smoothing length. This would suggest that, once again, the stability properties of SPH are improved by using kernels whose Fourier transform falls off more rapidly.

It is interesting to note that the standard implementation of viscosity has no effect on linear transverse waves. The stability properties of hexagonal lattices have also been considered in Morris 1994 with somewhat similar results. Three dimensional stability analysis also reveals unstable transverse modes which are reduced by the use of higher order spline interpolants. It can also be shown that the pressure differencing formulation does not exhibit these instabilities.

## 12. Conclusion

It is observed that the stability properties of SPH are improved by the use of kernels whose Fourier transforms fall off more rapidly. The instability first observed in exactly momentum conserving SPMHD in Phillips and Monaghan 1985 can be reproduced by a simple artificial equation of state (Eq. (4)). There are many applications which involve an equation of state having the effect of adjusting the background pressure acting between particles. Some examples are MHD, incompressible SPH, elastic-plastic flow and problems where we model a fluid experiencing an external pressure. The pressure adjustment, when SPH is implemented in an exactly momentum conserving form, influences the propagation of waves within the medium but this influence is reduced by the use of higher order spline interpolated kernels. Once the stress acting between particles changes sign, however, a different kernel must be used. It is possible to construct a kernel which varies with the stress acting between particles so as to ensure stability and realistic wave propagation. However, for complicated equations of state in two or three dimensions, this may not be feasible. Formulations of SPH which calculate the stress gradients by taking differences between the stress at neighbouring particles are observed to simulate the propagation of waves very well. The simulation of strong shocks, however, suffers since momentum is no longer conserved exactly. In many circumstances it may be best to split the stress tensor into a component which is always positive and evaluate the remainder using a differencing formulation. For example, in the case of MHD the magnetic pressure could be added to the hydrodynamic pressure and employed in an exactly momentum conserving form. This would permit the simulation of strong shocks for which pressure forces are dominant. It may also be possible to adjust the stress at each of the particles by a constant, so as to ensure it is positive everywhere, thus permitting exact momentum conservation. However, if the adjustment is extreme, a higher order spline interpolated kernel my be required in order to avoid instabilities and strong dispersive effects. Boundary conditions may also become complicated.

For all two dimensional and three dimensional applications using an exactly momentum conserving form, the use of kernels with compact support introduces instabilities in transverse modes on rectangular and hexagonal lattices of particles. The growth rates of these instabilities are observed to decrease dramatically as higher order spline approximations to the Gaussian are employed. These instabilities are not exhibited by differencing formulations.

In general, kernels more closely approximating a Gaussian will give better results, but will cost more computationally as the number of contributing neighbours increases. However, this cost may be offset since such kernels may permit a decrease in $h$. It is important to have an understanding of the detail one wishes to resolve in a given problem. This understanding combined with sound knowledge of the stability and accuracy of the method employed allows us to have confidence in the results obtained.